\documentclass[aps,pra,twocolumn,twocolumn,superscriptaddress]{revtex4-1} 
\usepackage{amsmath}
\usepackage{latexsym}
\usepackage{graphicx}
\usepackage{amsfonts}
\usepackage{braket}
\usepackage{natbib}
\usepackage{amssymb}
\usepackage{fancyhdr}
\usepackage[utf8]{inputenc}
\usepackage{colortbl}
\usepackage{xcolor}
\usepackage{subfigure}
\usepackage{psfrag}
\usepackage[colorlinks=true,citecolor=blue,linkcolor=blue]{hyperref}
\usepackage{tikz}
\usepackage{ulem}
\usepackage{bbold}
\usepackage{bbm}
\allowdisplaybreaks
\newcommand{\ope}[1]{\ensuremath{\hat{\text{#1}}}}
\newcommand\bbone{\ensuremath{\mathbbm{1}}}
\usetikzlibrary{arrows,shapes}

\newcommand{\Trace}[1]{\text{Tr}\left[#1\right]}		% Trace
\begin{document}

\title{Hybrid benchmarking of arbitrary quantum gates}
\author{Tobias Chasseur}
\affiliation{Theoretical Physics, Saarland University, 66123 Saarbr\"ucken, Germany}
\author{Daniel M. Reich}
\altaffiliation[Present address: ]{Department of Physics and Astronomy, Aarhus University, 8000 Aarhus C, Denmark}
\affiliation{Theoretische Physik, Universit\"at Kassel, Heinrich-Plett-Str. 40, 34132 Kassel, Germany}
\author{Christiane P. Koch}
\affiliation{Theoretische Physik, Universit\"at Kassel, Heinrich-Plett-Str. 40, 34132 Kassel, Germany}
\author{Frank K. Wilhelm}
\affiliation{Theoretical Physics, Saarland University, 66123 Saarbr\"ucken, Germany}

\begin{abstract}
We present a protocol for Interleaved Randomized Benchmarking of
arbitrary quantum gates using Monte Carlo sampling of quantum
states. It is generally applicable, including 
non-Clifford gates while  preserving key advantages of Randomized Benchmarking such as error amplification as well as 
independence from state preparation and measurement errors. This property is crucial for implementations in many contemporary systems.
Although the protocol scales exponentially in the number of qubits, it is
superior to direct Monte Carlo sampling of the average gate fidelity in both the total number of experiments by orders of magnitude and
savings in classical preprocessing, that are exponential. 
\end{abstract}

\maketitle

A central goal of quantum information science is to engineer a physical system capable of functioning as a scalable quantum computer, in order
to systematically outperform classical computers in certain applications. To this end,
it is imperative to drive arbitrary unitary evolution in a suitable quantum system consisting of  
$n\gg 1$ qubits and 
to benchmark the implementation of that operation.

Efficient benchmarking protocols, i.e., protocols that scale at most polynomially in $n$, are available for quantum operations in the
Clifford group \cite{Knill2008,Magesan2011,Flammia2011,Silva2011}, 
an important subset of quantum operations \cite{Nielsen2000,Dankert2009}. Monte Carlo sampling also applies to non Clifford gates but experimental and classical resources scale exponentially in that case. On the other hand Randomized Benchmarking (RB) is experimentally viable and therefore widely used in current experimental settings due to its independence from state preparation and measurement (SPAM) errors. 
In particular, randomized benchmarking (RB) is a method to estimate the average error of the Clifford group based on the fidelity of random Clifford gate sequences
\cite{Knill2008,Magesan2011}. This remarkable construction hinges on the Clifford group forming a unitary 2-design for 
$\mathcal{SU}(d=2^n)$. Another key prerequisite for the scalability
of RB is that $\mathcal{C}$ can be simulated efficiently on a
classical computer, based on the Gottesman--Knill theorem
\cite{Gottesman1999}. However, for the same reason, 
quantum algorithms based on those Clifford gates alone cannot outperform a
classical computer. To realize 
the full potential of quantum computation, one has to access the full
unitary group which is generated by $\mathcal{C}$ and one additional
non-Clifford gate, e.g., a single qubit gate such as the $\pi/8$
gate. Moreover, the Solovay-Kitaev theorem states
that every gate can be implemented to arbitrary precision by a sequence of elements of this set
with length logarithmic in the accuracy \cite{Dawson2006}. However said length scales exponentially in the number of qubits
\cite{Barenco1995,Knill1995} and therefore this is only efficiently applicable to few qubit gates. While some algorithms are known to achieve quantum speedup with a polynomial number of two-qubit gates, a universal quantum computer requires the capability to directly perform dense unitaries, i.e, gates that cannot be constructed by a polynomial number of one- and two-qubit gates. Accordingly it is necessary to benchmark
such arbitrary quantum gates as efficiently and reliably as possible.

Monte Carlo sampling of the average gate fidelity 
allows for the validation of arbitrary quantum gates \cite{Flammia2011,Silva2011,Reich2013}.
It requires significantly less resources than the canonical approach, which is to
extract this information from full quantum process tomography (QPT)
but is 
limited by SPAM. Its 
scaling in 
both experimental and classical resources, 
although favorable compared to QPT, is still exponential in $n$. SPAM 
typically includes single Clifford gates since standard  
candidates for quantum computing only prepare the ground state and
measure in the Pauli Z-basis \cite{Nielsen2000}. More general
initial states and a complete measurement basis can be  realized
via Clifford transformations. 
The fidelity of specific gates of the Clifford group can be estimated
with interleaved randomized benchmarking (IRB) \cite{Magesan2012}; 
its restriction to Clifford gates is not fundamental but the
generalization to arbitrary gates is highly challenging \cite{Chasseur2015}: 
Simulation and inversion of the sequence becomes increasingly
  difficult as alternating Clifford and e.g. $\pi/8$ gates generate
  the full special unitary group. Here, we show how arbitrary gates can be benchmarked by replacing the
inverting gate at the end of each IRB sequence with Monte Carlo
sampling of the resulting quantum state. 
Our approach preserves major advantages of RB such as independence of
SPAM and 
amplification of small errors via long sequences. In addition it outperforms 
direct Monte Carlo sampling of the average gate fidelity 
in the number of measurements 
and yields an exponential saving in classical computational
resources.

We first briefly review the original RB protocol
\cite{Knill2008,Magesan2011} as well as IRB
\cite{Magesan2012,Chasseur2015}. 
RB provides an estimate for the average fidelity of a unitary 2-design
such as the Clifford group based on the idea that random sequences of
Clifford gates also randomize the effect of error channels. 
For every sequence of $y$ Clifford gates $\ope{C}_j$, $1\leq j\leq y$, there
is a unique $\ope{C}_{y+1}$ inverting the sequence which can be efficiently
found via the Gottesmann--Knill theorem. Therefore, by applying the
sequence and its inverse to an initial state $\hat{\rho}_0$
and measuring the survival probability of that state, the
sequence fidelity is accessible experimentally. Averaging over all
possible sequences  and making the additional assumption of a gate
independent error channel $\Lambda$ 
results in an average sequence fidelity
\begin{align}
  \Phi_y = \frac{1}{{\sharp \mathcal{C}}^y}\sum_{\{C_j\}\in{\mathcal{C}}^y} \Trace{\hat{\rho}_0 \left(C_{y+1}\prod_{j=y}^1 (\Lambda C_j)\right) (\hat{\rho}_0) }\text{,}\label{eqn:del1}
\end{align}
where $\sharp$ denotes cardinality and 
$C(\rho)=\ope{C}\hat{\rho}\ope{C}^\dagger$ is the operation of the Clifford gate. 
The reverse order of
the product ensures the correct arrangement of the gates with the earlier operation applied to the state appearing on the right of the latter operation.

Using that $\mathcal{C}$ is a group, this can be rewritten as
$\Phi_y\equiv \Trace{\hat{\rho}_0 \Lambda_{\rm{twirl}}^{y}(\hat{\rho}_0)}$ where 
$\Lambda_{\rm{twirl}}$ -- the twirl of the error channel $\Lambda$ over the Clifford group \cite{Dankert2009} --  is completely
depolarizing, i.e., $\Lambda_{\rm  twirl}(\hat{\rho})=p\hat{\rho}+\frac{1-p}{d}\bbone$ with decay parameter $p$ \cite{Gambetta2012}. The average fidelity associated with the channel $\Lambda_{\rm twirl}$ then is $\Phi=p+\frac{1-p}{d}$; the average sequence fidelity becomes
\begin{align}
\Phi_y=\Trace{\hat{\rho}_0\left(p^y\hat{\rho}_0+\frac{1-p^y}{d}\bbone\right)}
=\frac{d-1}{d}p^y+\frac{1}{d}\label{eqn:noPMfid}\text{.}
\end{align}
To access $p$, one has to estimate $\Phi_y$ for several sequence lengths $y$ by sampling over a small subset of possible sequences for each $y$; $p$ and hence $\Phi$ are derived by fitting the experimental data to
an exponential decay. Incorporating the error channels for SPAM and $C_{y+1}$ leads to $\Phi_y=Ap^y+B$ which leaves the exponential decay unchanged and therefore yields a protocol robust against imperfect SPAM \cite{Knill2008,Magesan2011}. An important extension to RB is the IRB protocol that sets limits to the fidelity of a single Clifford gate $\ope{V}_\mathcal{C}$ using the fidelity of this gate interleaved with a random sequence, i.e., of the combined error channel $\Lambda_V\Lambda_\mathcal{C}$, in comparison with the fidelity obtained for the Clifford group \cite{Magesan2012}. This assessment of specific gates not only provides information of possible error sources but can be used directly for model free optimal control in the experiment \cite{Egger2014,Kelly2014}.

Potential loopholes in RB and IRB such as gate dependent errors and leakage can be accounted for via considering linear maps acting on linear maps on density matrices instead of just linear maps on density matrices \cite{Chasseur2015}. Specifically, the extension of the sequence length by one acts as a linear map $T$  on the operator on $\hat{\rho}$ representing the shorter sequence (which depends on $V$ for IRB). The average sequence fidelity is a linear functional of the $y$-fold product of $T$ resulting in a multi--exponential fidelity decay, 
\begin{equation}
\Phi_y=\sum_ia_i\lambda_i^y\,.  \label{eqn:model}  
\end{equation}
The eigenvalues, $\lambda_i$,  of the linear operator $T$ are close to real and their absolute values are smaller than or equal to one. The resulting fidelity decay can be fitted  using just a handful of different exponential decays \cite{Chasseur2015}.

The technique used in Ref. \cite{Chasseur2015}
does not depend on $V_\mathcal{C}$ being an element of the Clifford
group. Considering a general gate $V$ outside the Clifford group, the
unitary matrix representing an ideal implementation of the sequence
can be  quite general since $\mathcal{C}$ and $V$ generate a dense
subset of the whole special unitary group.
The same holds for the inverse $C_{y+1}$. 
Its construction therefore would be highly challenging and defeat 
the concept of constructing quantum computing from of a restricted set of gates. 
                              
Alternatively, one could be content to approximate the inverting gate using the Solovay-Kitaev theorem, which states that any gate can be composed out of a small number $l$ of gates
 depending only logarithmically on the permitted inaccuracy but exponentially on the number of qubits \cite{Barenco1995,Knill1995}. 
                              RB protocols require to neglect the error rate $\varepsilon_{y+1}$
 associated with $C_{y+1}$, i.e.,
$\varepsilon_{y+1}$ should be much smaller than the error of the
sequence. Since both sequence and inverting gate are composed of the
same gate set, this is roughly equivalent to $l$ being much smaller
than $y$. Satisfying  this is possible only for  $\varepsilon_V$ and
$\varepsilon_\mathcal{C}$ sufficiently small so that $y$ is large while
the Hilbert space dimension must be kept small as it enters the sequence length exponentially in
 the Solovay-Kitaev algorithm. In other words, satisfying  $l_i\ll
y$ implies the ability to implement an arbitrary quantum gate to a
relatively high precision, i.e., availability of a universal quantum
computer.

Given that these two related ideas do not work, we present a more practical approach to the problem.
Consider the fidelity of a specific sequence $\textbf{y}$, written as a vector of gates.
\begin{align}
\Phi_{\textbf{y}}&= \Trace{\rho_0 C_{y+1}\prod_{j=y}^1 (V \Lambda_V
  \Lambda_j C_j)(\rho_0)} \notag\\
&\equiv \Trace{\rho_{\rm id}^{\textbf{y}}\rho_{\rm act}^{\textbf{y}}}\text{,}\label{eqn:mc1}
\end{align}
where $\rho_{\rm
  id}^{\textbf{y}}={C_{y+1}}^{-1}(\rho_0)=(\prod_{j=y}^1C_j)(\rho_0)$
is the state ideally generated by the sequence and determined on a
classical computer, and  
$\rho_{\rm act}^{\textbf{y}}$
the one actually realized by applying the gates $V$ and $C_j$
(including their errors $\Lambda_V$ and $\Lambda_j$) in the experiment. 
Equation (\ref{eqn:mc1}) is of the form used in
Refs. \cite{Flammia2011,Silva2011,Reich2013} to estimate the overlap
of two states via \emph{Monte Carlo sampling}. Employing the notation of  Ref. \cite{Reich2013}, the states are
rewritten in the bases of 
the generalized Pauli matrices on $n$ qubits normalized for the
canonical scalar product defined by the (unnormalized) trace, $\mathcal{W}=\frac{1}{\sqrt{d}}\mathcal{P}^{\otimes n}$,
\begin{align}
\Phi_{\textbf{y}}&=\Trace{\rho_{\rm id}^{\textbf{y}}\rho_{\rm act}^{\textbf{y}}}=
\sum_k^{d^2} \Trace{W_k \rho_{\rm id}}\Trace{W_k \rho_{\rm act}}\notag\\
&\equiv \sum_k^{d^2} \chi_{\rm id}(k)\chi_{\rm act}(k)\notag
=\sum_k^{d^2} {{\chi_{\rm id}(k)}^2}\frac{\chi_{\rm act}(k)}{\chi_{\rm id}(k)}\\
&\equiv \sum_k^{d^2} \text{Pr}(k) X_k\,,\label{eqn:distribution}
\end{align}
where Pr$(k)=\chi_{\rm id}(k)^2$ and $X_k=\frac{\chi_{\rm
    act}(k)}{\chi_{\rm id}(k)}$.  $\sum_k^{d^2} \text{Pr}(k) =1$
since $\sum_k {\chi_{\rm id}(k)}^2=\Trace{\rho_{\rm id}^2}$ and
$\rho_{\rm id}$ being a pure state; 
and therefore $\text{Pr}(k)$ can be used as a sampling probability where the expectation value of the corresponding 
sampling is the desired fidelity $\Phi_{\textbf{y}}$.

This is the core of our
approach: Instead of actually implementing the gate that inverts the
random sequence and measuring the error on identity, we treat
$\Phi_{\textbf{y}}$ as a state fidelity which is estimated with Monte
Carlo sampling. Following Eq. \eqref{eqn:distribution}, 
this consists in choosing 
a total of $L$ Pauli measurement operators $W_{k_l}\in \mathcal{W}$,
$1\leq l\leq L$, according to the
sampling probability Pr$(k)$ and measure it (and hence $X_{k_l}$)
$N_l$ times. 
We summarize the IRB protocol with Monte Carlo sampling
of quantum states 
as follows:
\begin{enumerate}\setlength{\itemsep}{0pt}

\item For the characterization of a single arbitrary quantum gate $V$ we rely on IRB, hence we use RB to assess the average fidelity of all Clifford gates to set a reference point.
\item We choose $q$ different values of $y$ such that the sequence fidelities $\Phi_y$ can be assumed to provide a reliable fit; this  means the $\Phi_y$ shall be close to neither one nor the fidelity limit for long sequences.
\item For each of these values, we 
choose $m$ different sequences $\textbf{y}$ of random Clifford gates
interleaved with the gate $V$.
They are used to estimate the average fidelity $\Phi_y$ by comparing
the actual and ideal state,
cf. Eq. \eqref{eqn:mc1}, via Monte Carlo sampling.
\item To this end,
it is necessary to determine 
the ideal
state on a classical computer, i.e., to  multiply $2y$ unitary
matrices onto 
the pure initial state vector. This cannot be done efficiently since $V$ is not necessarily a Clifford gate; hence it scales as $\mathcal{O}(yd^2)$.
\item We choose $L$ measurement operators $W_k$ at random, following the
distribution Pr$(k)$ defined in Eq. (\ref{eqn:distribution}).
\item For each of those we apply the sequence \textbf{y} and measure the operator $W_k$ which we repeat $N_l$ times.
\item We determine an estimate for the sequence fidelities $\Phi_y$ by
averaging over all $N_l$ measurements, the $X_k$ for all $L$
measurement operators $W_{k_l}$, and the $m$ different sequences as given by Eq. (\ref{eqn:distribution}).  
\item We fit $\Phi_y$ to the multi--exponential decay,
  Eq. (\ref{eqn:model}), analogously to the original IRB, 
and derive the combined average error as $\frac{\Phi_{y=1}}{\Phi_{y=0}}$.
\item  We calculate the error rate of the arbitrary $n$ qubit gate $V$ as
$\varepsilon_V=\varepsilon_{\mathcal{C} \times
  V}-\varepsilon_{\mathcal{C}}$ and 
estimate the lower and upper bounds as
$\text{max}\left(0~,\left(\sqrt{\varepsilon_{\mathcal{C} \times
        V}}-\sqrt{\varepsilon_{\mathcal{C}}}\right)\right)^2 $ 
and $\left(\sqrt{\varepsilon_{\mathcal{C} \times
      V}}+\sqrt{\varepsilon_{\mathcal{C}}}\right)^2$ as in the
original IRB. 
\end{enumerate}

The parameters of the protocol are chosen as follows: 
A valid fidelity estimation via RB requires sufficient experimental
data for a fit to a (multi--)exponential decay; hence $q$ different
values for $y$, all provided with a substantiated estimate for
$\Phi_y$. Because it is sufficient to fit to only a handful exponential decays $q$ can be chosen relatively small.
The amount $m$ of different sequences for each value of $y$ has been
parameter independently upper bounded using the leading order in gate errors \cite{Wallman2014,Chasseur2015} yielding $m$ not larger than $100$, which can be directly transferred to our case. 
Higher order corrections to the uncertainty originating from finite $m$ can be bounded using the fact that $\Phi_y$ lies in the range $[0,1]$ and invoke Hoeffding's inequality \cite{Magesan2012b}.
We choose sequence lengths $y$ in a way that the error is neither too 
small to be measured efficiently nor so big that the decaying terms are already close to zero. 
This condition is satisfied for 
\begin{align}
\varepsilon y=\mathcal{O}(1)\text{,}\label{eqn:range}
\end{align}
as can easily be seen using the simplified model of a single decay.

In Monte Carlo sampling, there are two sources for inaccurate fidelity
assessment, namely the sampling inaccuracy due to \emph{a)} the incomplete subset of the measurement operators
and \emph{b)} that due to the finite number of measurements. The 
inaccuracies can be bounded by Chebyshev's and Hoeffding's inequality,
respectively to be allowed to exceed $\frac{\alpha}{2}$ with a probability of at most
$\frac{\delta}{2}$. Given an acceptable error bound, this leads to an
estimate of the total number of experiments.

The sampling inaccuracy \emph{a)} is bounded by  Chebyshev's inequality which
provides an upper limit to the probability of deviating from the mean
value of a distribution, depending on its standard deviation, 
\begin{align}
 \text{Pr}\left(\left|Z-\left[ Z\right]\right|\geq\frac{\sigma_Z}{\sqrt{\delta}}\right)\leq \delta\text{.}
\end{align}
Here, 
$Z\equiv\frac{1}{L}\sum_{l=1}X_{k_l}$ is the fidelity estimate
obtained by the random choice of measurement operators $W_{k_l}$ and $[Z]$ its
classical expectation value, i.e.,  $\Phi_{\textbf{y}}$. The variance can be estimated as
\begin{align}
\sigma_Z^2&=[Z^2] - [Z]^2 
=\sum_{l=1}^{L}\sum_{k_l}\text{Pr}\left(\frac{X_{k_l}}{L}\right)^2-\Phi_{\textbf{y}}^2\\
&\leq \frac{1}{L} \sum_k \chi_{\rm act}{(k)}^2 =\frac{1}{L} \Trace{\rho_{\rm act}^2}\leq \frac{1}{L}\text{,}\notag
\end{align}
using the fact that $\rho_{\rm act}$ is a density matrix but not
necessarily pure. Thus
\begin{align} \text{Pr}\left[\left|Z-\Phi_{\textbf{y}}\right|\geq\sqrt{\frac{2}{L\delta}}\right]\leq \frac{\delta}{2}\text{,}
\end{align}
where the outer brackets denote the ceiling function, 
and the choice $L=\left\lceil8/(\alpha^2 \delta)\right\rceil$ ensures the intended inequality.
To limit the deviation \emph{b)} due to a finite number of measurements one
relies on Hoeffding's inequality, 
\begin{align}
 \text{Pr}\left(\left|S-\left\langle S\right\rangle\right|\geq\alpha/2\right)\leq 2\text{exp}\left(-\frac{\alpha^2}{2\sum_i(b_i-a_i)^2}\right)\text{.}
\end{align}
$S$ is the sum over random variables $Y_i$ with outcomes in the range
$[a_i,b_i]$, here the adequately normalized sum of all $\sum_l N_l$
single shot measurements, and $\langle S\rangle=Z$. Since 
the measurement outcomes of Pauli matrices are bimodal, they 
are situated at the boundaries of the respective range $[a_i,b_i]$. 
Therefore, the range over variance ratio is most suitable for
Hoeffding's inequality. To ensure that the probability to exceed
$\frac{\alpha}{2}$ is at most $\frac{\delta}{2}$, it suffices to
demand 
\begin{align}
 \frac{\delta}{2}&\overset{!}{\geq}2\text{exp}\left(-\frac{\alpha^2}{2\sum_l 4N_ld^{-1}\left(LN_l\chi_{\rm id}(k)\right)^{-2}}\right)\text{,}\\
 \intertext{which, with the natural choice $N_l \propto {\chi_{\rm id}(k)}^{-2}$, is satisfied for}
 N_l&=\left\lceil\frac{8}{dL\alpha^2{\chi_{\rm id}(k)}^2}\text{log}\left(\frac{4}{\delta}\right)\right\rceil\text{.}
\end{align}
Compared to Refs. \cite{Flammia2011,Silva2011,Reich2013}, the total
inaccuracy $\alpha$ as well as the probability $\delta$ of exceeding it were
chosen smaller by a factor of two to simplify the further treatment. 

The classical average over the total number of experiments can be
estimated as follows: 
\begin{align}
[N_{\rm exp}]&=\sum_{l=1}^L\sum_{k_l=1}^{d^2}\text{Pr}(k_l) N_{k_l}\notag\\
&\leq \sum_{l=1}^L\sum_{k_l=1}^{d^2}\left(\text{Pr}(k_l)+\frac{8}{dL\alpha^2}\text{log}\left(\frac{4}{\delta}\right)\right)\notag\\
&=L\left(1+\frac{8d}{L\alpha^2}\text{log}\left(\frac{4}{\delta}\right)\right)
\notag\\
&\leq 1+\frac{8}{\alpha^2\delta}+\frac{8d}{\alpha^2}\text{log}\left(\frac{4}{\delta}\right)\,.\label{eq:Nexp}
\end{align}
Equation \eqref{eq:Nexp}
is also valid for direct Monte Carlo sampling of the average gate fidelity
and represents an exponential speedup compared to 
full QPT which scales as  $\mathcal{O}(d^4)$ \cite{Reich2013}. An important aspect is the scaling with 
$\frac{1}{\alpha^2}$. It is key to the advantageous scaling of IRB with Monte Carlo sampling
of quantum states in comparison with direct Monte Carlo sampling of
the average fidelity as shown below. 

For the resource estimate, we aim that the inaccuracy of fidelity measurements should be one order of
magnitude smaller than the error rate $\varepsilon$. Average gate
fidelities are not fundamental quantities of physics but estimators on
how good a quantum algorithm composed of a set of gates
performs. Therefore any attempt at an overly precise characterization of gate errors does not yield a valuable gain in  
information. In addition,  
the systematic uncertainty $\alpha_{\text{\tiny IRB}}$ of IRB caused
by Clifford gate errors limits the accuracy that can reasonably be
achieved; even more so for other methods not robust against SPAM
errors. Based on Eq. (\ref{eqn:range}),
$\Phi_y\sim1-y\varepsilon$ such that uncertainties in its estimation affect the estimate of $\varepsilon$ roughly with a factor of $\frac{1}{y}$. 
Therefore relative errors in $\Phi_y$ approximately translate to relative
errors in $\varepsilon$. 
With the above statement and that the inaccuracy of a IRB based estimation $\alpha_{\text{\tiny IRB}}$ is aimed to be $\alpha_{\text{\tiny IRB}}\propto
\varepsilon$ one chooses an inaccuracy $\alpha_{\text{\tiny MC}}(y)$ for the Monte Carlo sampling of sequence fidelities $\Phi_y$ resulting in an estimation not much more precise than $\alpha_{\text{\tiny IRB}}$; it scales linearly with $\varepsilon y$ which is in the order of $1$ . Therefore
$\alpha_{\text{\tiny MC}}(y)$ varies distinctly but not excessively over the $q$ different sequence
lengths $y$  but depends on neither the error rate $\varepsilon$ nor the Hilbert
space dimension $d=2^n$. For the sake of simplicity, let $\alpha_{\text{\tiny MC}}$ be defined as an effective average value for 
$\alpha_{\text{\tiny MC}}(y)$ setting an average on to what precision each sequence fidelity has to be assessed. $\alpha_{\text{\tiny MC}}$ as a system independent constant 
of the protocol, can safely be assumed to not deceed $10^{-1.5}$.

The above derivation of $\alpha_{\text{\tiny MC}}(y)$ ensures the required accuracy for each of the $q\times m$ single sequence fidelity rather than just for the resulting estimate for $\varepsilon$. This provides a reasonable fit to the decay function as each data point provides sufficient accuracy. Exploiting that in a more rigorous way may result in an improvement of prefactors but can not improve the scaling since $q$ and $m$ are largely system independent \cite{Wallman2014,Chasseur2015}.

The total number of experiments then sums up to
\begin{align}
  \left[N_{\rm exp}\right]\leq qm\left(1+\frac{8}{\alpha_{\text{\tiny MC}}^2\delta}+\frac{8d}{\alpha_{\text{\tiny MC}}^2}\text{log}\left(\frac{4}{\delta}\right)\right)\text{,}
\end{align}
which differs by a factor of $qm
\frac{\alpha^2}{\alpha_{\text{\tiny{MC}}}^2}$ compared to direct
Monte Carlo sampling of the average fidelity \cite{Reich2013}.
Translating this factor into numbers relating to 
recent advances in the implementation of quantum gates 
as well as the error threshold for quantum computing highlights the advantage of our protocol. A specific set of values taking into account recent experimental results \cite{Barends2014,Harty2014, Corcoles2013} corresponds to $q=20$, $m=50$ and
$\varepsilon=10^{-3}$ based on relatively high error rates of two
qubit gates. These
values yield $\alpha=10^{-4}$ and two orders of magnitude of improvement in the total number of experiments via the above factor. 

Another concern regarding scalability is the use of classical computational resources. 
Although more easily accessible, classical resources are not infinite
and therefore become relevant eventually, especially for 
Monte Carlo sampling where classical resources scale exponentially with a higher exponent than the number of experiments. 
The sampling of measurement operators can be done using conditional
probabilities, 
scaling with $n^2 d^2$ for states and $n^2d^4$ for processes and hence outperforming the naive approach of calculating all Pr$(k)$
\cite{Silva2011, Reich2013}. Accounting also for 
the necessity to calculate $\rho_{\rm id}$ for each sequence, the classical resources needed for our protocol scale as
\begin{align}
 N_{\rm class}=\mathcal{O}\left(\frac{qm}{\alpha_{\text{\tiny MC}}^2}\left(\frac{d^2}{\varepsilon}+n^2d^2\right)\right)\text{,}
\end{align}
compared to $\mathcal{O}\left(\frac{1}{\alpha^2}n^2d^4\right)$ for
direct Monte Carlo sampling of the average gate fidelity. Hence, we
obtain an exponential speedup of $\mathcal{O}\left(d^2\right)$ in classical resources in addition to the reduction of the number of experiments.

Combining the currently best but individually restricted methods for estimating quantum
fidelities--interleaved randomized benchmarking and Monte Carlo
sampling--we have extended the former to arbitrary quantum
operations, outside of the Clifford group, while avoiding the enormous
overheads and SPAM dependence associated with the latter. The extension to non-Clifford
gates is made possible by treating the RB sequence fidelity as a state
fidelity that can be estimated with Monte Carlo sampling. This avoids
the actual accurate physical implementation of the inverting gate in the RB
sequence which, for a non-Clifford gate, would require availability of
a universal quantum computer. Our protocol inherits from IRB
robustness with respect to SPAM errors, which -- for current
experimental settings -- can completely mask the actual error
channel. As a conclusion the resulting hybrid algorithm is a viable tool for SPAM--independent, robust benchmarking of arbitrary quantum
gates. While non-exponential scaling is beyond reach -- and might well be impossible -- the proposed protocol reduces the total number of
experiments compared to direct Monte Carlo sampling of the gate fidelity due to error amplification and yields exponential savings in
the classical preprocessing resources.

We acknowledge travel support from the EU through QUAINT as well as funding through SCALEQIT and a Google Faculty Research Award. 

\bibliography{LRB.bib}
\bibliographystyle{apsrev4-1}
\end{document}